# Transport evidence for Fermi-arc mediated chirality transfer in the Dirac semimetal Cd$_3$As$_2$

**Authors:** Philip J.W. Moll[1,2,*], Nityan L. Nair[1], Toni Helm[1,2], Andrew C. Potter[1], Itamar Kimchi[1], Ashvin Vishwanath[1], and James G. Analytis[1,3,*]

**Affiliations:**

[1]Department of Physics, University of California, Berkeley, California 94720, USA

[2]Max-Planck-Institute for Chemical Physics of Solids, Noethnitzer Strasse 40, D-01187 Dresden, Germany

[3]Materials Science Division, Lawrence Berkeley National Laboratory, Berkeley, California 94720, USA

*Correspondence to:   philip.moll@cpfs.mpg.de , analytis@berkeley.edu

**The dispersion of charge carriers in a metal is distinctly different from that of free electrons due to their interactions with the crystal lattice. These interactions may lead to quasiparticles mimicking the massless relativistic dynamics of high-energy particle physics[1–3], and they can twist the quantum phase of electrons into topologically nontrivial knots – producing protected surface states with anomalous electromagnetic properties[4–9]. These effects intertwine in materials known as Weyl semimetals, and their crystal symmetry protected analogs, Dirac semimetals[10]. The latter show a linear electronic dispersion in three dimensions described by two copies of the Weyl equation, a theoretical description of massless relativistic fermions. At the surface of a crystal, the broken translational symmetry creates topological surface states, so-called "Fermi Arcs"[11], which have no counterparts in high-energy physics nor conventional condensed matter systems. Here we present Shubnikov-de Haas oscillations in Focused Ion Beam prepared microstructures of Cd$_3$As$_2$ that are consistent with the theoretically predicted "Weyl orbits", a kind of cyclotron motion that weaves together Fermi arc and chiral bulk states[12]. In contrast to conventional cyclotron orbits, this motion is driven by the transfer of chirality from one Weyl node to another, rather than momentum transfer of the Lorentz force. Our observations provide evidence for direct access to the topological properties of charge in a transport experiment, a first step towards their potential application.**

The bulk electrons in topological semimetals are described by an ultra-relativistic dispersion relation, E(k)=±ℏv$_F$ **σ*k**, resembling the Weyl equation for massless spin-1/2 particles. Here σ is a pseudo-spin-1/2 degree of freedom that is energetically locked parallel or anti-parallel to the momentum, k, of the electron, giving electrons definite chirality k ±σ. Applying electromagnetic fields to Weyl or Dirac semimetals induces a pumping of electric charge between Weyl nodes with opposite chirality, a phenomena known in high energy physics as the chiral anomaly[13–15]. At the surface of these materials, this anomalous chirality transfer is facilitated by topologically protected surface arcs, the so-called "Fermi arc" surface states, which act as a pipeline connecting opposite chirality Weyl points[11,16]. Recently, $Na_3Bi$[17] and $Cd_3As_2$[18] have been predicted to be three-dimensional bulk Dirac semimetals. This has sparked significant research interest and the linear dispersion in these materials was confirmed by ARPES[19–24] and STM[25] experiments. A number of unusual material properties such as strong linear magnetoresistance[19,26] and high mobilities[27] were identified that are potentially linked to the relativistic nature of the Dirac quasiparticles. Yet the prospect of studying ultra-relativistic particles and their accompanying topological surface states, as well as potential applications exploiting their unusual behavior naturally requires a more direct measurement capable of revealing both the relativistic dynamics and topological surface states. The main aim of this study is to present such evidence in four-terminal transport measurements, showing the possibility of detecting and manipulating chiral states in Dirac semimetals.

One intriguing fingerprint of Weyl quasiparticles in the electronic transport properties in strong magnetic fields has been recently predicted by Potter, Kimchi and Vishwanath (PKV)[12]. This "Weyl orbit" (Fig.1) weaves together the chiral states in the bulk with the topological Fermi-arc states on opposite surfaces into a closed orbit. Its quantization produces a distinctive contribution to the quantum oscillation spectrum that provides an observable signature of the chiral and topological character of these materials. This closed orbit is strikingly different from typical electrons orbiting around a Fermi surface in a metal as the quasiparticle experiences zero Lorentz force on the chiral path segments traversing the bulk.

The main result of this study is an additional quantum oscillation frequency observed in microstructures smaller than the mean-free-path that exhibits characteristics of both surface-like and bulk-like states, as naturally expected for Weyl orbits. The microstructures were prepared from $Cd_3As_2$ single crystals by FIB etching (Fig.1, see Methods). Down to the smallest thickness of L=150*nm*, the magnetoresistance at temperatures below *100K* shows pronounced Shubnikov-de Haas oscillations signaling the low effective mass of the charge carriers and the high crystal quality of the devices (Fig.1b). Quantum oscillations on bulk crystals have reported one single frequency[19,26], arising from an essentially spherical 3D Fermi surface in agreement with ARPES and STM experiments. This single bulk frequency ($F_B$) is also consistently observed in all of the studied parent bulk crystals as well as all microstructures. However, when the field is applied perpendicular to the [0 1 0] surface (0°), a second frequency $F_S$= 61.5*T* appears which is distinct from the higher harmonics of the bulk. We find an effective mass of $0.044m_e$ for the bulk, similar to previously reported measurements on bulk crystals[26], and a similar mass of $0.050m_e$ in the additional orbit $F_S$ (See methods section).

The value of the observed surface frequency is compatible with the prediction for Weyl orbits. Their trajectory combines segments of both chiral bulk and Fermi arc surface states, and the quantum oscillation frequency $F_S$ can be estimated from the time spent in each of them[12]:

$$F_S = E_F k_0 / e \pi v_F \approx 56T$$

The length of the surface Fermi arc in reciprocal space may be approximated as $k_0 \approx 0.8$ $nm^{-1}$ using the k-space separation of the Dirac points determined by ref.[19]. The bulk quantum oscillation frequency and effective mass provide direct access to the Fermi energy $E_F = 192 meV$ and the Fermi velocity $v_F = 8.8 \ 10^5$ $m/s$. The resulting estimate of $F_S \sim 56T$ is in good quantitative agreement with the measured $F_S = 61.5T$.

The quantum oscillation frequency dependence on the angle between the magnetic field and the surface normal shows a clear signature of a surface state as shown in Figure 2, where 0° denotes fields perpendicular to the surface and 90° parallel to it. The angular dependencies of both frequencies are strikingly different: while the low frequency $F_B$ (blue) remains essentially constant and is observed at all angles, the second frequency $F_S$ (red) strongly increases as the field is tilted away from the surface. $F_S(\theta)$ is well described by a $\cos(\theta)^{-1}$ dependence, indicating that the field component perpendicular to the surface is relevant for the orbit. This angle dependence is a hallmark of two-dimensional (2D) Fermi surfaces. The bulk band structure however does not support a 2D Fermi surface and it is not observed in our bulk crystals.

While the angle dependence clearly suggests a surface character of the quantum path associated with the additional frequency, the quantum oscillations also show pronounced bulk-like characteristics which are unexpected for orbits simply consisting of surface states. First of all, the additional quantum oscillation is only observed in samples where the bulk mean-free-path is longer than or comparable to the sample thickness. The low-field transverse magnetoresistance for in-plane fields quantitatively confirms our devices to be in this limit. The resistance maximum at small fields is a hallmark signature of quasi-ballistic transport in clean metals known as the "Knudsen effect" (Fig.3a, see methods).

The frequency spectrum of the quantum oscillations is found to be strongly thickness dependent (Fig.3b). No trace of the surface frequency $F_S$ has been observed in devices thicker than $3\mu m$, in agreement with its absence in our bulk quantum oscillation measurements. As the sample thickness is reduced, its relative weight compared to the bulk frequency strongly increases, and devices thinner than $500nm$ are dominated by the surface oscillation. The increase in the various studied samples follows an exponential behavior, with an exponent of d = $675nm$. This value should be compared to the bulk mean free path estimated from transport as $l = v_F \tau = v_F \ m^* / (ne^2 \rho_0) = 1.0 \mu m$, where $\rho_0 = 55 \mu \Omega cm$ is the zero-field resistivity in our crystals at 2K, n = 2 $k_F^2$ / ($3\pi^2$) = 2.5 $10^{18}$ $cm^{-3}$ the bulk carrier density estimated for the two-fold degenerate spherical Fermi surface and $k_F = m^* v_F / \hbar = 3.3 \ 10^8 m^{-1}$ the Fermi momentum extracted from the Shubnikov-de Haas oscillations. This estimate of the mean free path is in good quantitative agreement with the observation of the Knudsen flow maximum. Intriguingly, this thickness dependence matches well with the expectation for Weyl orbits. Ordinarily, any scattering phases acquired suppress quantum oscillations in which case the quantum oscillations would be expected to decay exponentially in L over the quantum mean free path $l_Q$ which is generally much shorter than the mean-free path measured in transport. However, for large fields the bulk chiral Landau level is expected to exhibit extra resilience to such dephasing and allow for quantum oscillations for thicker samples. As the closed Weyl orbit contains two path segments traversing the bulk, each at opposite chirality, the effective bulk path length is twice the device thickness L. The agreement between $2d = 1.25 \mu m$ and $l = 1.0 \mu m$ suggests that the relevant thickness scale for the appearance of the additional quantum oscillations is comparable to the bulk transport mean-free path.

Additional clue as to the origin of the Shubnikov-de Haas oscillations comes from their phase as a function of field. We observe deviations from ideal periodicity of the oscillations in inverse magnetic field, appearing as a continuously drifting phase as shown in Figure 4. While similarly subtle deviations from periodicity can arise due to g-factor band splitting in strongly spin-orbit coupled materials[28], the direction of the shift is opposite to those previously observed in experiment: The peak positions $B_n$ are shifted towards higher fields compared to the purely periodic case, while spin-splitting should shift them towards lower fields. On the other hand, this direction and magnitude of shift of the Landau levels is expected for Weyl orbits. Non-adiabatic corrections are expected to appear due to field-induced tunneling between Fermi arc states and bulk states, occurring as the orbiting quasiparticles approach the Weyl node. A single Weyl orbit will encounter four such tunneling processes, leading to a phase deviation that is in quantitative agreement with what we observe (See Fig.4 and methods).

The discussion so-far considered the Dirac material $Cd_3As_2$ as two independent Weyl subsystems overlapping in k-space. However, magnetic fields can break the crystal symmetry protecting the superimposed Weyl nodes of opposite chirality and thus introduce a gap. In sufficiently strong fields, the particle approaching the node may thus tunnel into the oppositely dispersing Fermi-arc on the same surface instead into the bulk, thus forming closed orbits purely from Fermi-arc states similar to the surface states of topological insulators. While our experiments cannot rule out localized topological surface states as the origin of the oscillations, the thickness dependence of the surface state amplitudes and the non-adiabatic corrections are more suggestive of the Weyl orbit in the present field range. In both cases, the Fermi arcs constitute in the quantum orbit and our results confirm the detection of topological surface currents in the microstructures.

However quantum oscillations may also arise from trivial surface states or from the defect layer introduced by the fabrication technique, without the involvement of topological states. Thus it is essential to find ways to probe if topological states participate in the observed orbit, setting the involved orbit apart from topologically trivial states. The existence of a "saturation field" above which oscillations were expected to cease was predicted to provide such test[12]. However, a reexamination of the equation for quantum oscillations, reproduced in Eqn. 1 below, reveals that the oscillations associated with the Weyl orbit actually persist up a much larger field of order $F_S=56T$ in the present instance (see methods), and the thickness dependence of the saturation field is highly complex. Consequently, the Weyl surface oscillations are expected to persist over the full range of fields explored in this experiment, in agreement with our observations. Additional evidence for the topological nature of the surface oscillations comes from their remarkable resilience against surface disorder. We have purposely introduced significant surface damage by almost normal incidence FIB-irradiation. Remarkably, the surface state oscillations grow upon increasing disorder, in strong contrast to the usual disorder dampening of conventional surface oscillations. The counter-intuitive increase of amplitude with increasing surface damage, however, would be a natural consequence of the protection of topological surface states (see methods).

Another striking piece of evidence for the non-trivial nature of the quantum orbit comes from a distinguishing feature of the Fermi arc orbit: the quantum oscillation phase depends on the thickness of the sample, L, given by:

$$B_n^{-1} = e\, k_0^{-1}\, (n\, \pi v_F\, /\, E_F - L) \quad (1)$$

The position of the $n^{th}$ resistance minimum, $B_n$, is predicted to follow Eq.1, where $k_0$ describes the length of the Fermi arc surface state in k-space, $v_F$ is the Fermi velocity, and $E_F$ the Fermi energy.

To test this concept, we fabricate a sample geometry that is by design sensitive to such a thickness dependence of the phase. In Figure 5 we show a sample structured with a triangular cross-section as well as a rectangular one as a reference. The quantum oscillations for fields perpendicular to the surface for each device (0° for the rectangle, 60° for the equilateral triangle) are strikingly different. While the rectangular device clearly shows the presence of a surface state, the triangular one shows only the bulk frequency without any sign of the second surface frequency. Crucially, all surfaces were fabricated under the exact same conditions from the same piece of crystal. The sample was tilted with respect to the ion beam to ensure a grazing incidence condition for every surface, both on the rectangular and the triangular device. Both devices have comparable cross-section and surface area by design, so that a trivial surface state acting as a parallel conductance channel should lead to observable surface quantum oscillations in both of them. The absence of the surface state in the triangle is unexpected for oscillations arising from a trivial surface state, yet a natural consequence of Weyl orbits. Unlike conventional surface states that are confined to a single surface, the Weyl orbit cannot be observed in triangular geometries: All paths are of different length, each contributing to the field induced density-of-state modulation at different magnetic field $1/B_n$ (see Eq.1, Fig.5). This results in destructive interference due to a sum of oscillations with random phases, rendering the quantum oscillations unobservable in experiment. In contrast, all quantum paths involve bulk path segments of the same length L in a rectangular geometry, and thus all contribute to a density-of-state modulation at the same field. This directly evidences that the orbit associated with the frequency $F_S$ is sensitive to the shape and size of the bulk underneath the surface, in contrast to those arising from trivial surface states.

The ensemble of presented results highlights an essential aspect of the additional quantum oscillations appearing when Dirac fermions are confined into microstructures smaller than the mean free path: They share characteristic features of both surface-like and bulk-like oscillations. Such hybrid characteristics arise naturally from the idea of the mixing of chiral bulk- and Fermi-arc surface-states into a coherent orbit. Nevertheless, there remain questions that challenge our current understanding of topological matter on microscopic length scales and in strong magnetic fields. For example, understanding the exact amplitude of the Weyl oscillations may require the extension of the present theory into the quantum limit, where the number n of occupied Landau levels cannot be treated as being large. Furthermore the understanding of the mixing of chiral states in strong magnetic fields and the influence on the Weyl orbit should be included and experimentally investigated in the future. By utilizing FIB structuring, we have demonstrated an experimentally simple route towards studying strongly confined topological matter, which will both increase our understanding of the transport characteristics of Fermi-arc states as well as provide a path to investigate the potential of these materials in future electronic applications.

**Acknowledgments:** The Focused Ion Beam work was supported by the SCOPE-M center for electron microscopy at ETH Zurich, Switzerland. We thank Philippe Gasser, Joakim Reuteler and Bertram Batlogg for FIB support, and Maja Bachmann for performing magnetoresistance measurements. A.C.P. was supported by the Gordon and Betty Moore Foundation's EPiQS Initiative through Grant GBMF4307. We also thank S. Teat and K. Gagnon for their help in conducting X-ray diffraction measurements at the Advanced Light Source (ALS) beam line 11.3.1 and N. Tamura for micro-diffraction on beam line 12.3.2. N.T. and the ALS are supported by the Director, Office of Science, Office of Basic Energy Sciences, Materials Sciences Division, of the U.S. Department of Energy under Contract No. DE-AC02-05CH11231 at Lawrence Berkeley National Laboratory and University of California, Berkeley. Transport experiments, material synthesis and Focused Ion Beam microstructuring were supported by the Gordon and Betty Moore Foundation's EPiQS Initiative through Grant GBMF4374. Single-crystal x-ray refinements and theoretical support were funded by the Quantum Materials FWP, U.S. Department of Energy, Office of Basic Energy Sciences, Materials Sciences and Engineering Division, under Contract No. DE-AC02- 05CH11231.

**Author contributions:** P.J.W.M. microstructured the crystals, performed the measurements and data analysis. N.L.N. synthesized and characterized the single crystals. T.H. analyzed the crystal structure. I.K., A.C.P. and A.V. contributed the theoretical treatment. P.J.W.M and J.G.A. designed the experiment. All authors were involved in writing the manuscript.

# Figures

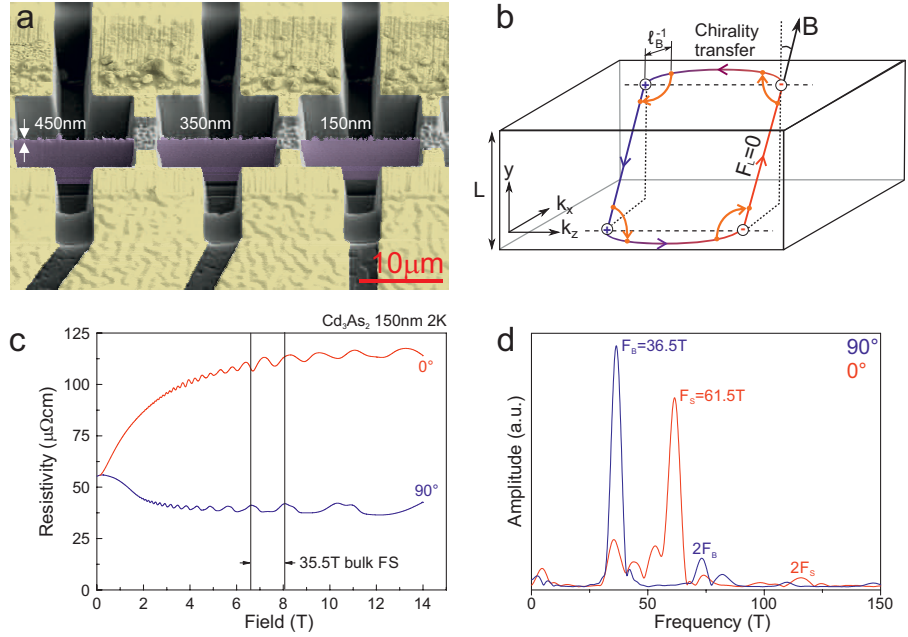

**Fig 1. Surface oscillations in microstructures**
  a) SEM micrograph of a typical sample prepared by FIB cutting. The active devices are $4\times10\mu m^2$ free-standing sheets (purple) of varying thickness connected to contact pads (yellow). The crystallographic direction perpendicular to the polished surface is [0 1 0] and parallel to it [1 0 0], which define the plane of rotation. This entire sample consists of one contiguous slice of a $Cd_3As_2$ crystal, ensuring that all devices are made from the exact same starting material and have the same orientation during the experiments.
  b) Sketch of the Weyl orbit in a thin slab of thickness L in a magnetic field B. The orbit involves both the Fermi arc surface states connecting the Weyl nodes of opposite chirality, and the bulk states of fixed chirality (blue and red). In strong magnetic fields, quasiparticles may tunnel through the energy barrier separating the bulk states from the surface over a distance associated with the magnetic length $l_B$ (orange arrows). Note that as the Weyl orbit mixes processes in real and reciprocal space, the (x,y) coordinates of this sketch are in reciprocal and the z coordinate is in real space.
  c) Magnetoresistance and (d) its Fourier transform measured on the thinnest device (150$nm$) at 2$K$ for both fields parallel (90°) and perpendicular (0°) to the surface. The main finding of this study is directly evident in the raw data: While parallel fields lead to a single frequency (incl. spin-splitting at higher fields), an additional higher frequency component $F_S$ associated with the surface oscillations appears for perpendicular fields. This high frequency is clearly distinguishable from higher harmonics of the low frequency $F_B$.

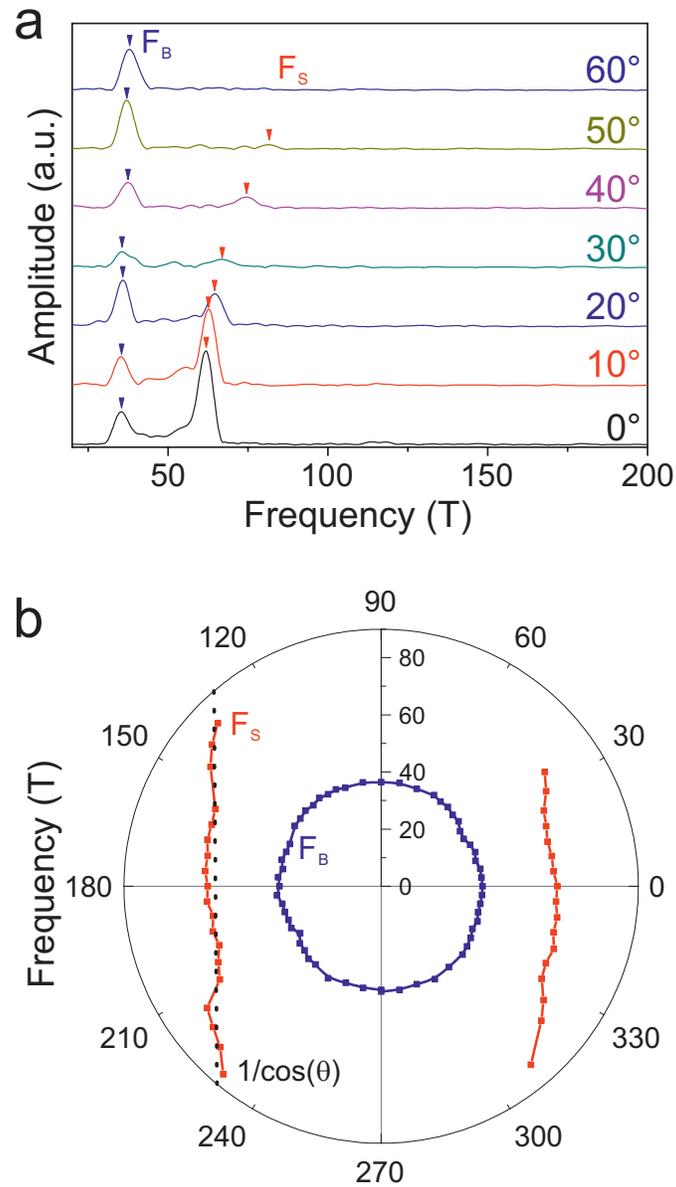

**Fig. 2. Angle dependent oscillations**

a) Angle dependence of the quantum oscillation spectrum measured in the 150*nm* device. As the field is rotated away from the perpendicular configuration (0°), the surface frequency $F_S$ increases and shrinks in amplitude, while the bulk frequency is unaffected.
b) Polar plot of the angle dependence for the bulk $F_B$ and the surface frequency $F_S$. $F_B$ is almost isotropic while $F_S$ increases as the field is rotated towards a parallel configuration. It follows a surface-like $\cos(\theta)^{-1}$-dependence resembled by a straight, off-center line in a polar plot.

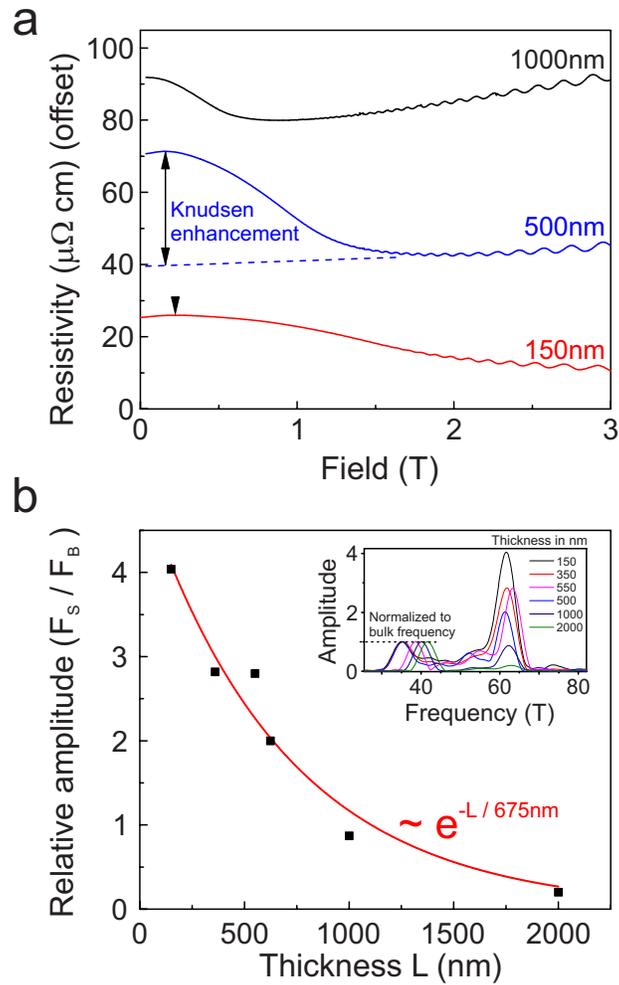

**Fig. 3. Thickness dependence**

a)  A magnetoresistance maximum was observed in all studied samples for fields applied parallel to the surface (90°). This peak arises from the semi-classical Knudsen effect observed in quasi-ballistic transport when the cyclotron radius becomes comparable to the sample dimensions.

b)  Relative amplitude of the surface oscillations compared to the bulk oscillations, for fields perpendicular to the surface at $2K$. The surface oscillations are unobservable for devices thicker than $3\mu m$, and their relative weight grows with thinning of the samples.

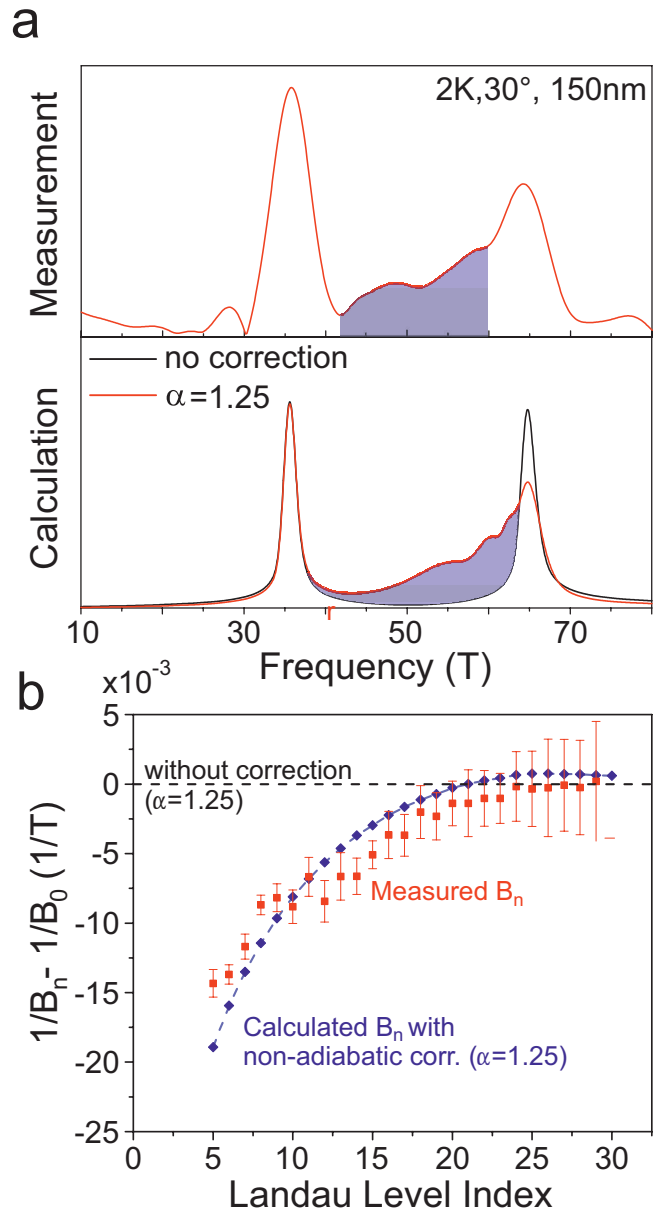

**Fig. 4. Non-adiabatic corrections**

a) Comparison of a measured oscillation spectrum to simulations, with and without the non-adiabatic correction term. Without non-adiabatic corrections, the expected spectra are well-defined peaks of similar broadening. The non-adiabatic corrections lead to an asymmetric reduction of the effective amplitude in high magnetic field, and shift spectral weight from the main surface peak to lower frequencies (shaded area).

b) Difference between the observed resistance maxima corresponding to the n-th landau level, $1/B_n$, and the positions $1/B_0$ extrapolated from low fields using the usual $1/B$ periodicity. The difference can be well explained by including the non-adiabatic correction term using $\alpha=1.25$ (blue dots).[12]

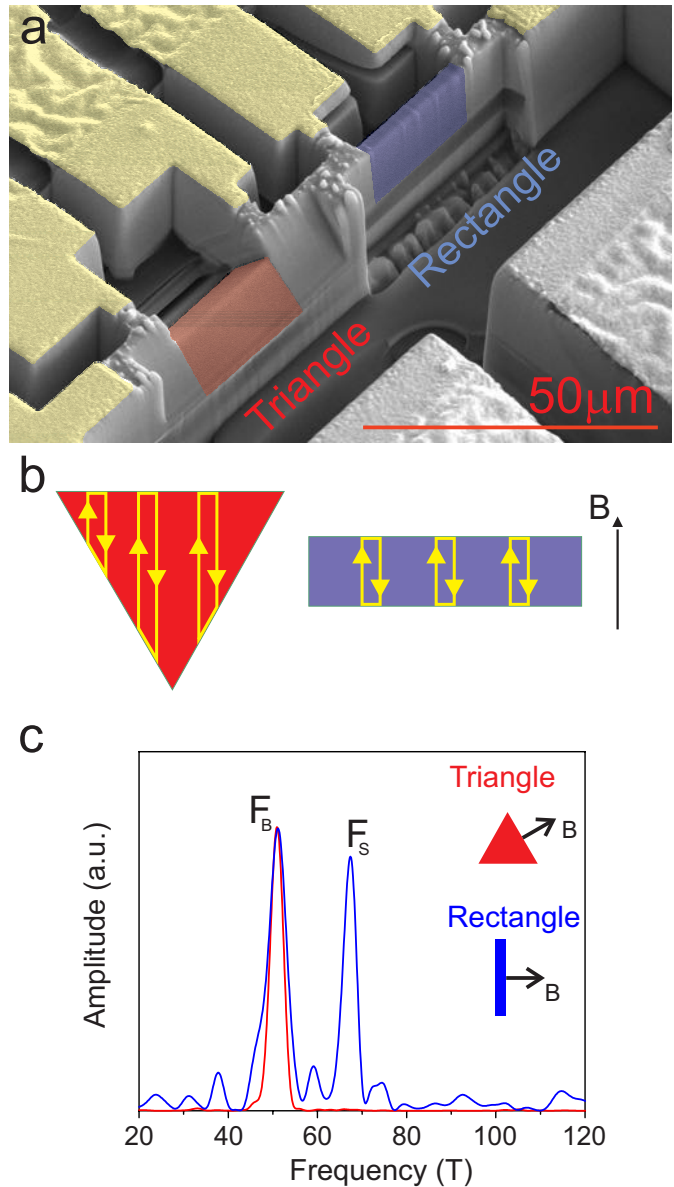

**Fig. 5. Absence of surface oscillations in triangular samples**

a) SEM image of triangular and rectangular devices. The rectangular sample is L = 0.8$\mu m$ wide, 3.2$\mu m$ tall and 5$\mu m$ long. The other device features an equilateral triangular cross-section with a base of a = 2.7$\mu m$. Both devices have a similar cross-sectional area (rectangle: 0.8 × 3.2 = 2.6$\mu m^2$, triangle: a=3.1$\mu m^2$) and surface length (rectangle: 2 × (3.2+0.8) = 8.0$\mu m$, triangle: 8.1$\mu m$). The crystallographic direction perpendicular to the surface of the rectangular device is [1 0 2] and [0 1 0] parallel to the surface, which defines the plane of rotation.

b) Sketch of the Weyl orbits for rectangular and triangular cross-sections.

c) Frequency spectrum of the triangular and rectangular samples, for field orientations perpendicular to each of the surfaces (0° for the rectangle and 60° for the triangle).

# Methods

**Focused Ion Beam sample preparation -** The Focused Ion Beam (FIB) has proven to be a powerful tool to fabricate crystalline microstructures of high quality, e.g. ref. [29–33], to study meso- and microscopic transport phenomena such as the Weyl-orbit quantum oscillations presented in this study. In this process, the microstructures are carved out of macroscopic crystals using the FIB. Starting from millimeter-sized flux-grown single crystals presented in Extended Data Figure 1, we use a $30kV$ $Ga^{2+}$ ion beam to cut the crystal. Depending on the ion flux required, the irradiation spot can be as small as a few nm, and thus structures can be fabricated with high precision. The beam is guided across a negative image of the desired structure, which defines the parts of the crystal to be removed. Electric contacts to the device can also be made in situ in the FIB. A platinum precursor gas can be introduced into the chamber, and upon interaction of the ion beam with the gas adsorbed to the surface a conductive, Pt-rich film can be grown connecting the microstructure electrically to external leads.

The fabrication occurs in two steps: In the first step, a cross-sectional slice is cut from the crystal. Typical dimensions of this slice are $100\mu m \times 15\mu m \times 3\text{-}4\mu m$. $Cd_3As_2$ is found to be sputtered very effectively by the ion beam and even large volumes can be sputter etched with low ion currents at a reasonable rate. This renders $Cd_3As_2$ an ideal material for FIB microstructuring and the low currents contribute to the high crystal quality of the fabricated devices. For the coarse patterning in this step, currents of $2.5nA$ were used. This slice is then removed ex-situ under an optical microscope and transferred to a $Si/SiO_2$ substrate with predefined gold leads connecting the center to bonding pads. The slice is glued in the center using a small drop of epoxy and cured for $1h$ at $100°C$. Good results were obtained with both Araldite Rapid and Stycast 1266 epoxies. $Cd_3As_2$ is found to react adversely with the platinum precursor gas leading to an insulating interface with non-ohmic behavior between the platinum contacts and the crystal slice. To avoid this barrier formation, the slice was sputter-coated in $100nm$ gold, which was later removed by FIB etching from the top-side of the studied microdevices. A more detailed description of the process was previously published in ref[34]. In the second step, the mounted slice is again introduced into the FIB for final patterning into the thin bar shape shown in the main text. Again the high sputter rate allowed us to use currents as low as $40pA$ for the structuring process. First, the coarse outline of the final structure was patterned at $300pA$, and it was checked that no stray connection due to redeposited material remained. Then the final thickness was adjusted by $40pA$ milling. The scan strategy for the final approach was a line-by-line, single pass mill from both sides of the device to ensure symmetric, highly polished surfaces. The strong bulk-quantum oscillations indicate the high crystallinity of our microstructures, which was additionally confirmed via X-Ray microdiffraction at the Advanced Light Source (Lawrence Berkeley National Laboratory).

**Nature of the surface layer -** FIB microstructuring is a very gentle process inducing very little bulk damage as evidenced by the strong bulk quantum oscillations present in our $Cd_3As_2$ devices. On the other hand, the impacting ions are known to damage the crystal surface and create a completely amorphous surface layer. Therefore particular care must be taken to investigate if a highly conductive layer is formed that exhibits conventional surface Shubnikov – de Haas oscillations. As discussed in the main text, the experimental evidence points against a trivial conductive surface layer as the source of the observed surface-like oscillations: (a) These heavily disordered damage layers are unlikely to show quantum oscillations due to their short mean-free-path, (b) it would not show adiabatic corrections to the quantum oscillation frequency and (c) the triangular device shows no sign of surface oscillations despite its comparable surface area and crystal orientation. Yet it remains essential to understand the surface layer, and therefore we analyze its expected thickness and nature by Monte Carlo simulations of ion interaction with $Cd_3As_2$, which provide an estimate of the expected surface layer thickness around $20nm$. The interaction of ions with matter can be well captured using Monte-Carlo simulations. We have performed a damage analysis including a calculation of the full cascade damage using the SRIM package (Stopping Range of Ions in Matter)[35]. Under realistic fabrication conditions of grazing incidence polishing ($88°$ to surface normal), we found a longitudinal stopping range of $7.6nm$ in $Cd_3As_2$. This parameter describes the mean implantation depth of Ga, with a standard deviation of $5.4nm$. At grazing incidence, the Ga-ion backscattering probability is strongly enhanced and thus the Ga-implantation is strongly suppressed compared to normal incidence. The main nature of the defects in the surface layers are cascade damages, resulting in a displacement of Cd and As atoms from their positions in the crystal lattice. The first $20nm$ of material surface are thus expected to be amorphous, with an exponential reduction of defects and damage deeper in the bulk. This is in agreement with the observation of quantum oscillations corresponding to the bulk frequency in our samples.

In the following we probe more directly if indeed topological surface states are involved in the surface-like orbit: One consequence of the topological character of the Fermi arc states involved in the Weyl orbit is their protection from surface disorder. This can be directly tested by purposely damaging the FIB-polished surfaces of $Cd_3As_2$. By irradiating one side of a device at close-to-normal beam incidence, the surface layer can be severely damaged and disordered. Upon increasing surface damage, the surface state quantum oscillations of a trivial surface state are expected to be dampened due to the decreased quasiparticle lifetime following the Lifshitz-Kosevich mechanism[36]. On the other hand, the topological surface state existing at the boundary between the periodic crystal bulk and the amorphous surface layer would remain unaffected, as it is simply pushed deeper into the bulk. To inflict maximal surface damage, exceptionally long beam dwell times at 30° angle between the beam and the surface normal were used to increase heating and cause deeper defect penetration (1*ms*, 80*pA*, 30*kV*). Under these conditions, a well visible pattern of shallow dimples appears on the damaged surface (Extended Data Figure (EDF) 2 b). The structure is of the same design as discussed in the main manuscript with a thickness of 810*nm* from the same $Cd_3As_2$ crystal, yielding a ratio of quantum oscillation amplitudes of bulk and surface states of 1.2 as expected at this thickness (Main figure 3). The same device was fabricated with polished surfaces and first well characterized in its pristine state. Then the same device was measured again after the irradiation, thus allowing to directly investigate the changes due to the irradiation.

The resistivity of the device increased by 55%, indicating the effective introduction of surface disorder due to the irradiation (EDF 2c). Remarkably, however, the amplitude of the Shubnikov-de Haas oscillations of both the surface and the bulk states have increased due to the irradiation (EDF 2d,e). This is in stark contrast to the expected strong reduction of amplitude due to disorder of a trivial surface state. However, this unusual result is a natural consequence of a topologically protected surface state. The current in such structures is expected to flow in three parallel paths: the crystal bulk, the topological surface state, and the disordered surface layer. By irradiating the surface at low incidence angle, the outermost damage layer becomes more disordered, as evidenced by the increase of the total device resistance. Therefore, the relative weight of the current flowing through the topological states and the bulk increases, leading to the unusual phenomenon of increasing surface state quantum oscillations upon increasing surface damage. This is further supported by the almost unchanged surface to bulk amplitude ratio due to irradiation (EDF 2e).

The robustness of the observed surface state quantum oscillations to artificial extensive surface damage strongly suggests a mechanism protecting the surface state from excess scattering at play, which would arise naturally from the topological properties of the Fermi arc states. This effect of topological protection from non-magnetic scattering has been shown in $Cd_3As_2$[20,22] as well as in topological insulators[6,7,37]. Clearly, such robust quantum coherent signatures are contrary to the expectations for trivial surface states.

**Quantum oscillation analysis -** Magnetic fields quantize the electronic degrees of freedom orthogonal to the field direction. The level spacing $\Delta E = \hbar\omega_c = B\ \hbar e/m^*$ grows linearly with increasing field strength B and the Landau levels are successively emptied as they are pushed above the Fermi level. This leads to a well-known modulation of the density-of-states at the Fermi level (DOS), and conversely to an oscillatory behavior of physical quantities depending on the DOS in high fields. A good overview of such quantum oscillatory effects is given in ref[36]. The effective mass m* is obtained from the temperature dependent dampening of the quantum oscillations due to thermal broadening of the Landau levels. The thermal dampening is described by the Lifshitz-Kosevich formalism and it can be shown that the oscillation amplitude follows the functional form $A(T)/A_0 = X/\sinh(X)$ where $X = 2\pi^2 k_B T / (\hbar\omega_c)$. Here $k_B$ denotes the Boltzmann constant and $\omega_c = eB/m^*$ is the cyclotron frequency. From fitting this temperature dependence to the measured amplitudes, as shown in EDF 3a, the effective mass is obtained as a fitting parameter as $0.044 m_e$ for the bulk. Using Onsager's relation, the frequency F of the quantum oscillations can be directly related to the extremal cross-sectional areas $S_k$ of the Fermi surface, as $F = 2\pi e\ S_k\ /\ \hbar$. As the Fermi surface of $Cd_3As_2$ is to good approximation spherical, $S_k = \pi k_F^2$, the Fermi wavevector can be estimated as $k_F = (4\pi eF/\hbar)^{1/2} \approx 3.33\ 10^7 m^{-1}$. The separation between the nodes of $0.8 nm^{-1}$ as reported in ref.[19] well exceeds the experimentally determined $k_F$. The Fermi velocity for our microstructure follows as $v_F = \hbar k_F/m^* \approx 8.76\ 10^5 m/s$. The Dirac equation now leads to an estimate of the Fermi energy above the Dirac point as $E_F = \hbar\ v_F\ k_F \approx 192 meV$.

Additional complications for the analysis of the Shubnikov-de Haas effect arise from the quasi-ballistic transport in our clean microstructures that are thinner than the bulk mean free path. At very low fields, a peak in the magnetoresistance with a broad maximum followed by a strong negative magnetoresistance is observed before eventually the bulk magnetoresistance is recovered at high fields (Main Fig. 3a). The quantum oscillations emerge ontop of this background. We note that this field configuration is transverse and thus no negative magnetoresistance associated with the chiral anomaly is expected. Instead this enhancement of scattering in ultra-pure systems is a semi-classical effect arising from the diffuse scattering of otherwise ballistic electrons at the boundaries of strongly confined microstructures[38]. It is well-studied in ultra-pure hydrodynamic systems, such as the viscous flow of $^3$He through capillaries (Knudsen-effect) or in high quality, geometrically confined conductors (Gurzhi-flow) such as semiconductor-heterostructures and clean metal whiskers [39]. The resistance maximum at the Knudsen peak occurs at maximal boundary scattering of the bent electron trajectories, i.e. at $2r_c = L$ up to a small numerical factor[40], where $r_c = \hbar k_F / (eB)$ is the cyclotron radius. As a result, the position of the maximum is expected to shift to higher fields as the sample thickness is decreased, as is observed in the $Cd_3As_2$ microstructures (See main Fig3a). The Knudsen effect is a direct consequence of ballistic electron motion between the opposite surfaces and thus directly evidences that the thickness of the studied microstructures is indeed comparable to the bulk mean free path.

**Non-adiabatic corrections** - The positions of the Landau levels associated with $F_S$ systematically deviate from the usual $1/B$ periodicity. Fig. 4 of the main text shows the deviation of each maximum of the resistance oscillation, $1/B_n$, from its expected position $1/B_0$ for a usual $1/B$ periodicity (blue dashed line). At low fields, the oscillations indeed are found to be periodic in $1/B$, yet slightly but consistently to deviate at higher fields. The deviations from periodicity are shown in Figure 4b.

We find quantitative agreement between the observed deviation and the expectations for non-adiabatic corrections of the Weyl orbit (Fig. 4b). Yet deviations of similar magnitude from periodicity could also occur from spin-splitting of the surface state alone, without the presence of Weyl orbits. A comparison of the presented data with the well-known predictions for spin-splitting, however, shows that the observed peak positions are qualitatively different from expectations in the spin-splitting scenario, further highlighting the unusual character of the observed oscillations (EDF 4). In the presence of strong magnetic fields, the Fermi surface volume for spin-up and spin-down electrons changes due to the Zeeman energy $\Delta E = \frac{1}{2}\mu_B g \sigma_z H_z$, where the z-axis is pointing along the magnetic field. This leads to an effective splitting of an initially spin-degenerate Fermi-surface at zero field (in the absence of spin-orbit coupling). In favorable cases, spin splitting can be directly observed as the appearance of two, well-separated peaks positioned symmetrically around the expected position in the absence of spin-splitting[41]. We do not observe split peaks in the surface state related oscillations. This could either be an indication for weak spin-splitting, or for a thermal or impurity broadened situation where direct splitting of the peaks is not observable despite strong spin-splitting. In the latter case, spin-splitting is known to modify the amplitude of the oscillations only, without affecting the phase of the oscillations[36]. Therefore spin-splitting of a trivial surface state alone is unlikely to explain the present data.

Another possibility may be the formation of a surface state akin to those in a topological insulator (TI). In this case, the spin-momentum locking on the Dirac cone changes the spin-splitting behavior and oscillation phase changes without any peak splitting are possible[42]. The reason for this is the opening of a gap in the Dirac spectrum due to the time reversal symmetry breaking of the applied magnetic field. This gap due to the Zeeman energy modifies the Landau level spectrum for electrons (at positive band filling): $E_N = v_F ( 2ne\hbar H + (g\mu_B H / (2v_F) )^2 )^{1/2}$

In the absence of spin-splitting (g=0), this reduces into the conventional square-root-dependence of Dirac systems. At low fields, i.e. large Landau level index n, this correction is negligible, yet becomes important at higher fields closer to the quantum limit (n=0). The resulting fields $B_n$, where a Landau level equals the Fermi Energy $E_F$ can be easily calculated using the material parameters self-consistently obtained from the quantum oscillation analysis (blue dots) and are contrasted to the measured peak positions (red dots), shown in EDF 4a.

Spin-splitting of a TI-like surface state is at odds with our observations for two reasons: (a) the expected deviation goes in the wrong direction and (b) the observed magnitude is incompatible with measured g-factors in $Cd_3As_2$. Spin-splitting would suggest that the fields $B_n$ occur at lower fields compared to the purely periodic case, they are on the contrary observed at higher fields. The shift to lower fields in the case of spin-splitting can be easily understood from the surplus Zeeman energy adding to the field dependence of the Landau levels in the absence of spin splitting (see figure below). This shifts the levels up in energy, and thus they intersect the constant chemical potential at lower fields, hence the upwards bending of the fields $B_n$ as shown above. In addition, a g-factor of 300 is required to obtain a shift of similar magnitude as observed by our experiment, which is an order of magnitude larger than the experimentally measured values[43].

On the other hand, such a downwards deviation as well as its magnitude is expected for Weyl orbits arising from non-adiabatic corrections: At small fields, the quasiparticle evolving on one of the surface states may only enter the bulk state at the Dirac point, where the gap between the surface and bulk states vanishes. In strong magnetic fields, however, the quasiparticle may tunnel through this band-gap into the bulk state before reaching the Dirac point over a distance given by the magnetic length $l_B = (\hbar / (eB))^{1/2}$. This shortens the effective path length on the surface (red arrows in Fig.1b) and thus leads to a non-adiabatic correction to the low-field surface frequency $F_{S,0}$ in strong fields given by

$$F_S(B) = F_{S,0} - 4 \alpha F_{S,0} l_B / k_0 \qquad (2)$$

, where α represents a material dependent parameter encoding the tunneling barrier between the surface and the bulk bands, and $F_{S,0}$ the field-independent surface frequency in the absence of strong non-adiabatic effects observed at low fields. The four tunneling processes involved in each Weyl orbit thus lower the effective quantum oscillation frequency $F_S$ in strong magnetic fields.

Including this non-adiabatic correction term well explains the observed deviation (Fig.4b, red points) from the pure 1/B periodicity in high magnetic fields. A fit to the observed data (blue points) yields a coupling parameter α~1.25, in good agreement with the theoretical predicted value of order unity[12]. This non-adiabatic correction to the $F_S(B)$ is also directly evident in the Fourier spectra: While the signature of the bulk frequency is a sharp, symmetrical peak centered at $F_B$, the surface frequency $F_S$ is asymmetric with significant spectral weight shifted to lower frequencies. This again is a result of the effective lower frequency in high magnetic fields, and can be well reproduced by a simple calculation inserting the field-dependent frequency given by Eq.2 into the Lifshitz-Kosevich formalism and Fourier-transform the resulting wave-form (Fig.4a of the main text). Unlike the spin-splitting driven phase modification, the non-adiabatic correction mechanism leads to a slowing-down of the oscillations at higher fields due to the effective reduction of flux enclosed in the quantum path that is cut short by the tunneling process. This naturally leads to the down-bending observed in our experiment.

**Thickness dependence of the oscillation phase -** Equation 1 of the main manuscript describes the expected width dependence of the oscillations arising from the Weyl orbit. Crucially, the thickness L does not affect the oscillation frequency, as the oscillation period Δ is independent of it: $\Delta = B_n^{-1} - B_{n-1}^{-1} = e\, k_0^{-1}\, \pi\, v_F / E_F$. In agreement with this expectation, no thickness-dependence of the frequency is observed. Instead it is expected to only affect the *phase* offset of the peaks vs 1/B and indeed the data shows no thickness dependence to the peak positions. However, experimentally detecting this phase shift is a challenge for four main reasons: 1) the surface- and bulk- frequencies are close to each other leading to significant beating which complicates the identification of the peak positions corresponding to the surface frequency, 2) the peak positions show an additional field dependence due to the non-adiabatic corrections, 3) the oscillatory signal is on top of a large non-linear and thickness-dependent background magneto-resistance due to finite size effects which must be subtracted to find the peak positions, and 4) the relative amplitude of the surface oscillations compared to the bulk oscillations by itself is strongly thickness dependent, which causes a thickness dependence of the beating pattern. Combined, these uncertainties in fitting the peak positions are larger than the separation between peaks, and therefore from the present data it is impossible to make a quantitative claim about the systematic dependence of the peaks. These difficulties are naturally avoided in the triangular geometry and thickness dependence experiments. By design, the triangular sample averages over parts of different effective width, leading to a cancellation of the surface signal as expected from Eq.1.

**Upper field limit for surface state oscillations -** Reference[12] predicted that quantum oscillations associated with the Weyl orbit persist only up to a maximum field that depends on sample thickness, $B_{sat} \sim k_0 / (eL)$. By repeating this analysis, we show that the thickness dependence is actually more involved. Instead, the surface arc oscillations persist up to a field that depends in a complicated non-monotonic fashion on film thickness, but which is at least as large as the quantum limit associated with the k-space area enclosed by the Fermi arcs. To see this, note that from Equation (1) above, we can see that the smallest integer, n=N, for which (1) has a solution is given by: $N = \left\lceil \frac{k_F L}{\pi} - \gamma \right\rceil$, where $k_F = E_F/v$, and $\lceil x \rceil$ indicates the closest integer to x that is larger than x. N represents the index of the last quantum level associated with the Weyl orbits that can be pushed across the Fermi-energy. The field at which level N crosses the Fermi-energy, i.e. the upper limit field beyond which Weyl surface-arc oscillations cease, is then: $B_{sat} = \frac{k_F k_0}{\pi} \frac{1}{N - \left(\frac{k_F L}{\pi} - \gamma\right)}$. The first factor in $B_{sat}$ is precisely the 1/B frequency of the surface state, $F_S = 56T$. The second factor depends strongly on L in a complicated oscillatory fashion, but is typically of order 1 (though possibly much larger when ($k_F L / \pi$) - γ is accidentally close to an integer value). Hence, we see that, in contrast to the claims of ref.[12], the upper limit field for observing Weyl surface arc oscillations is comparable to the quantum limit of the surface state, and is largely independent of sample thickness. In the present measurements, $B_{sat}$ is expected to be at least ~60T, and is hence unobservable in any non-pulsed field experiment.

**Raw oscillation data: 150nm device** - As quantum oscillations are phenomena most easily understood in Fourier space, we limit the discussion in the main manuscript on the analysis of FFT spectra. Fourier transforms, however, can be misleading and thus it is very important that the main results are clearly apparent in the raw data themselves. This is clearly the case: EDF 5 and 6 shows the oscillatory magnetoresistance after subtracting a $2^{nd}$ order polynomial to remove the non-oscillatory components. The same notation of angle is used as in the main manuscript. Scans were performed in 5° steps around the surface state orientations (0°,180°) and in 10° steps elsewhere for the 150nm device, and in 5° steps for the triangle. No averaging or symmetrization of the data between field sweeps of opposite polarity was performed, i.e. between sweeps with 180° angle difference. From the dataset EDF 5, the main Figure 2b showing the surface character of the oscillations was constructed. The main result of the manuscript can be seen as follows: For in-plane fields, i.e. 90°, 270°, a single frequency corresponding to the bulk frequency is observed. As discussed in the main text, both the observation of only one single frequency as well as its amplitude is in agreement with all other published data of quantum oscillations on $Cd_3As_2$. For fields perpendicular to the surface, i.e. at 0°, 180°, the spectrum is profoundly different. An additional oscillation of higher frequency is mixed with the bulk frequency, leading to a beating pattern. The second frequency gradually disappears as the field is rotated away from 0°, and is identified with the surface state oscillation from the FFT spectrum as described in the main text. Main Figure 5 showing the differences between the rectangular and triangular devices was constructed from the raw dataset shown in EDF 6. Also here the main points can be well seen without resorting to Fourier transforms: Despite its identical surface and cross-sectional area, the triangle does not show any sign of a second frequency at any angle, as can be seen from the angle-independent positions of the quantum oscillation peaks. In the rectangular case, however, strong beating appears when the field is perpendicular to the surface, signaling the second frequency. Also the characteristic angle dependence of surface states as presented in main Figure 2 can be clearly seen when one traces the maxima from the surface to higher angles.

## Additional References

# Extended Data Figures

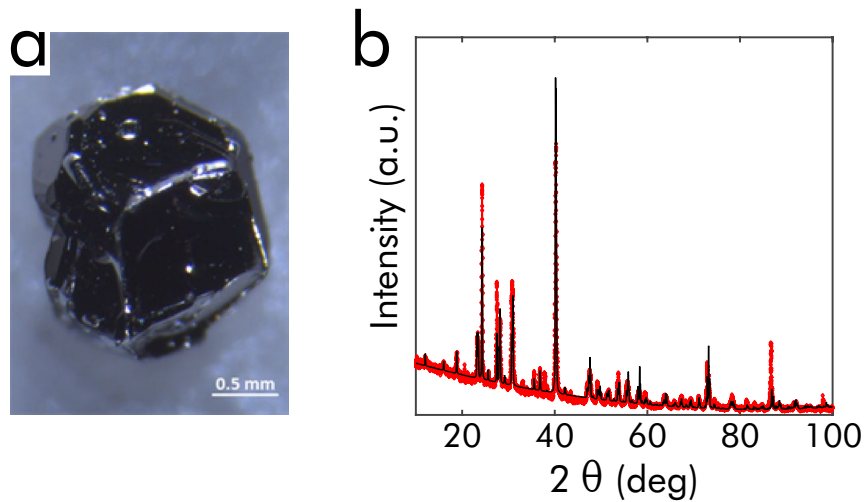

**Extended Data Figure 1: Crystal synthesis & characterization**
$Cd_3As_2$ single-crystals were produced using the flux growth technique with a 5:1 ratio of Cd flux to $Cd_3As_2$. Elemental Cd and As were placed into an aluminum oxide crucible with a quartz wool plug and sealed into a quartz ampule under vacuum. The ampule was heated to $825°C$ and was held for two days to ensure a fully homogenized mixture. It was then cooled at a rate of $6°C$ per hour to $425°C$, where it was centrifuged to remove excess Cd flux. Bulk crystals were found trapped in the quartz wool following this procedure, and were confirmed to be $Cd_3As_2$ by both power X-ray diffraction (PXRD) and single-crystal X-ray diffraction. The PXRD data is explained by the $I4_1/acd$ low-temperature phase of $Cd_3As_2$[44] and no parasitic phases were observed. Panel (a) shows a typical facetted crystal and (b) the PXRD spectrum.

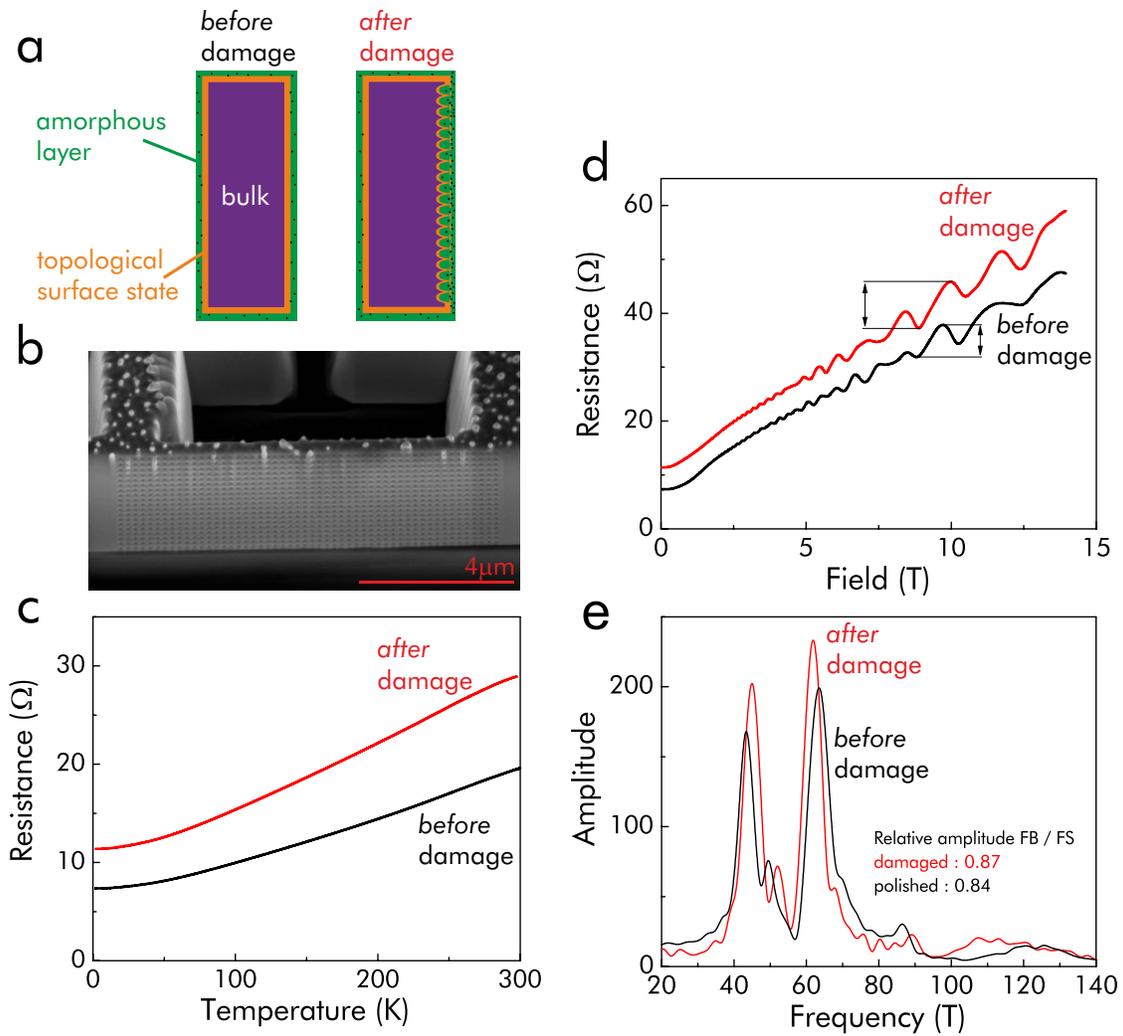

**Extended Data Figure 2**: **Surface quantum oscillations are resistant to intentional surface damage** a) Sketch of the device cross-section before and after the irradiation. Three possible current paths exist in these devices: The crystal bulk (purple), the topological surface states (orange), and the amorphous FIB-induced damage shell (green). b) Device after the heavy irradiation damage. The dimples due to the beam center impact are well visible across the whole device. The polished backside remained undisturbed by this procedure. c) Comparison of the same device with pristine surfaces and after damaging the surface. The resistance increases after irradiation, as expected for the increased scattering. d) Magnetoresistance of the $Cd_3As_2$ microstructure before and after irradiation; and e) the Fourier components of the quantum oscillations.

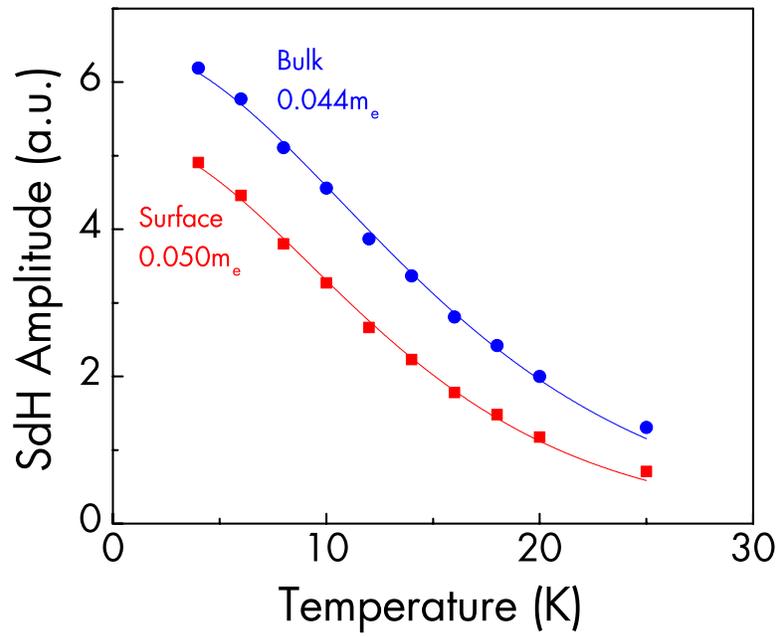

**Extended Data Figure 3: Effective mass analysis of bulk and surface oscillations** The temperature dependence of the amplitudes yields an effective electron mass in the bulk of $0.044m_e$, in good agreement to previously reported values [26]. The surface state appears slightly heavier ($0.05m_e$).

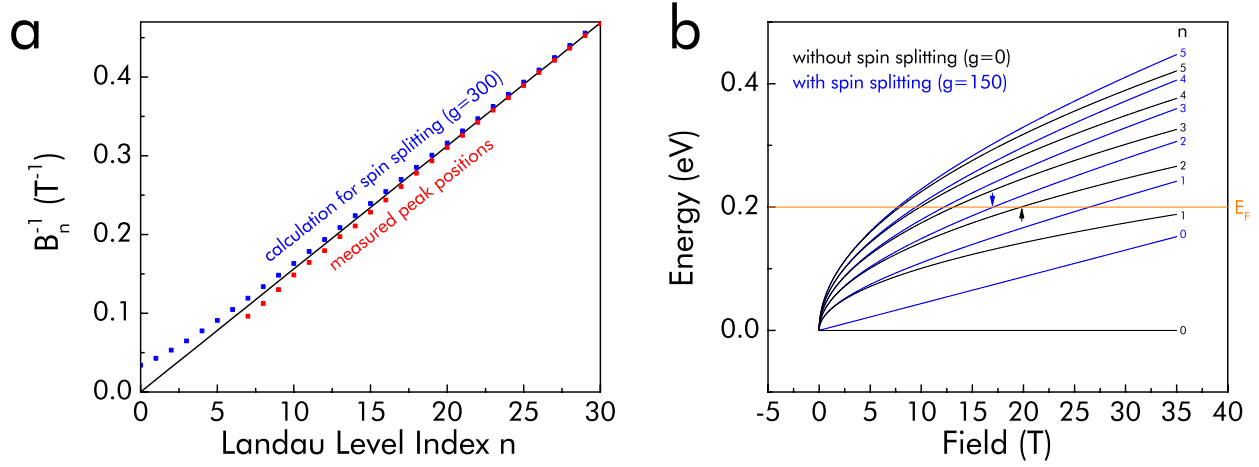

**Extended Data Figure 4: Spin splitting**: (a) Landau level positions calculated for spin-splitting (blue) and experimentally observed (red). The observed deviation is opposite to the expectation of spin-splitting, yet qualitatively and quantitatively consistent with the non-adiabatic corrections of the Weyl orbit process. (b) Landau fan diagram for Dirac systems.

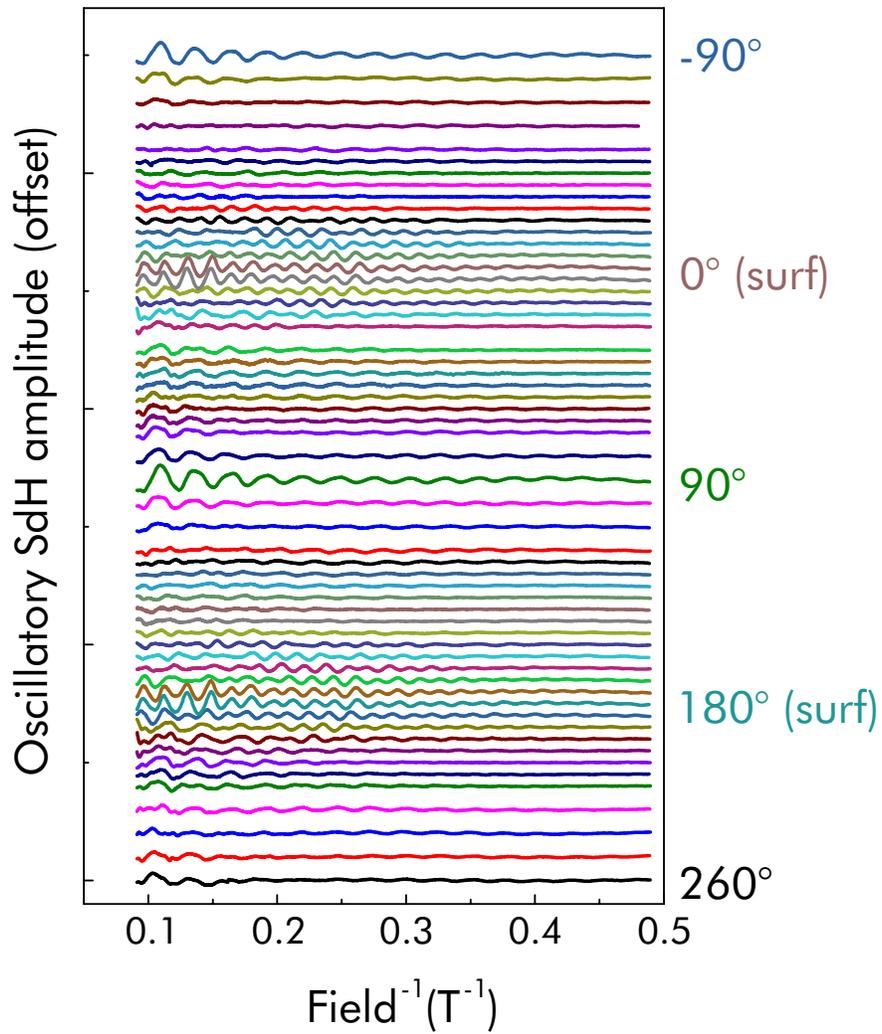

**Extended Data Figure 5: Raw oscillatory signal of the 150nm device**: Shubnikov-de Haas oscillations of the 150*nm* wide rectangular device, the smallest microstructure fabricated in this study. The appearance of the surface frequency can be well seen in the raw data as strong beating appears around 0°. Also the characteristic $\cos(\theta)^{-1}$ angle dependence of a surface frequency can be easily seen by following the peak positions to higher angles.

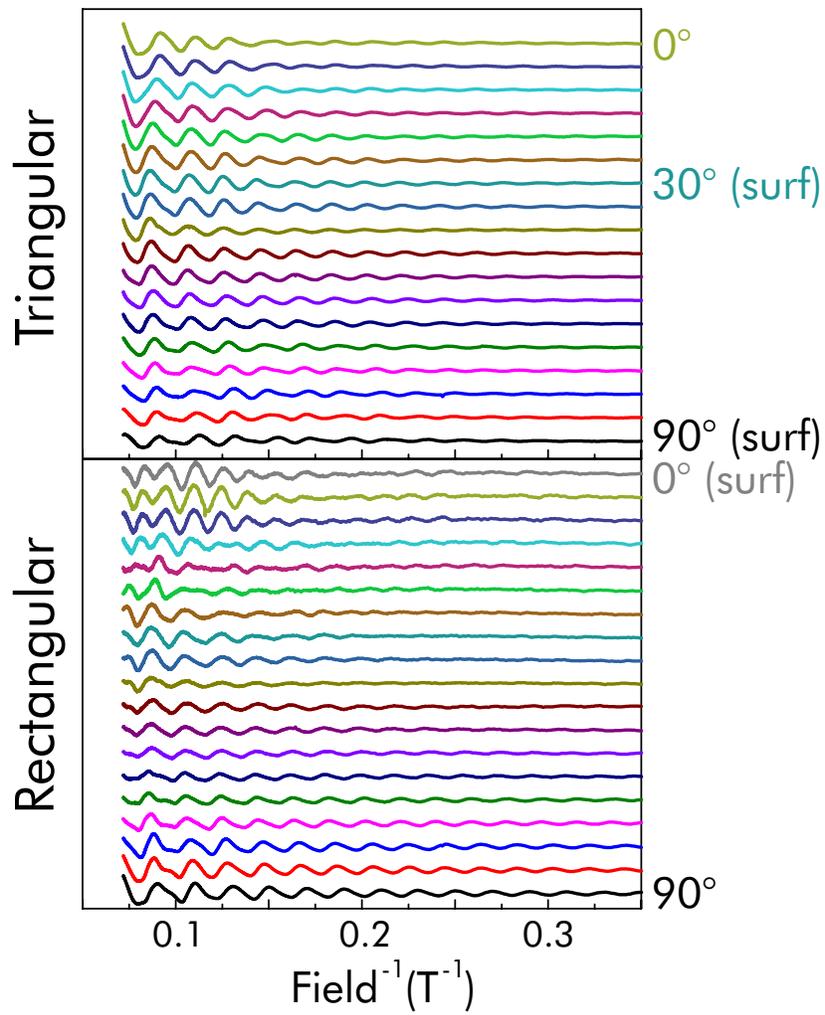

**Extended Data Figure 6: Raw oscillatory signal of the triangle/rectangle device**: Shubnikov-de Haas oscillations of the rectangular and triangular devices. While the rectangular device shows the characteristic beating, the peak positions in the triangle remain at the same fields.